\documentclass[showpacs,twocolumn,amsmath,amssymb,prb,superscriptaddress]{revtex4-1}

\usepackage{graphicx}
\usepackage{amsmath,amsfonts,amssymb,amsthm}
\usepackage{fancyhdr}
\usepackage{mathrsfs}

\usepackage[latin1]{inputenc}
\usepackage[colorlinks]{hyperref}
\begin{document}

\title{          Fully superconducting Bloch-oscillating transistor:
                 Amplification and bifurcation based on Bloch oscillations and counterflowing Cooper pairs           }

\author{J. Lepp\"akangas}
\affiliation{                    
Department of Microtechnology and Nanoscience, MC2, Chalmers University of Technology, SE-41296 G\"oteborg, Sweden
}
\affiliation{
       Department of Physics, University of Oulu, FI-90014 Oulu, Finland}

\author{A. Puska}
\author{L. \"Ak\"aslompolo}
\affiliation{Low Temperature Laboratory, O.V. Lounasmaa Laboratory, Aalto University, FI-00076 AALTO, Finland}

\author{E. Thuneberg}
\affiliation{
       Department of Physics, University of Oulu, FI-90014 Oulu, Finland}

\author{P. J. Hakonen}
\affiliation{Low Temperature Laboratory, O.V. Lounasmaa Laboratory, Aalto University, FI-00076 AALTO, Finland}


 \begin{abstract}
The Bloch-oscillating transistor (BOT) is an amplifier that utilizes semiclassical dynamics of 
states in energy bands under traveling quasimomentum.
In a BOT, a single quasiparticle tunneling across a base tunnel junction switches the  state
of a superconducting tunnel junction to a lower Bloch band, triggering a series of resonant Cooper-pair tunnelings through an emitter Josephson junction (Bloch oscillations). 
Here, we investigate experimentally and theoretically an alternative realization of this device, based only on superconducting tunnel junctions.
We discover new amplification schemes, where the periodic motion of the quasimomentum is used to control
charge transport between the electrodes.
Remarkable, in operation the resonant Cooper-pair transport across the base Josephson junction
occurs repeatedly to two opposite directions.
 \end{abstract}

 \maketitle

 \section{Introduction}
 
\begin{figure}[bt]
\includegraphics[width=\linewidth]{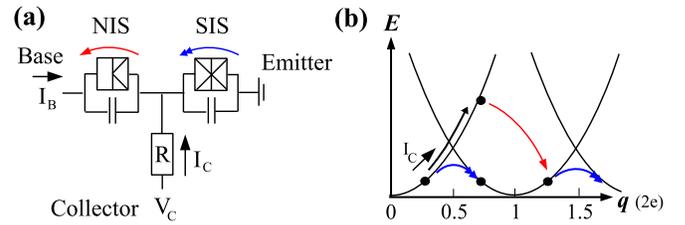}
\caption{
(a): The circuit diagram of a  Bloch oscillating transistor (BOT).
    The  collector is connected to the island through a large resistance ($R\gg R_Q$) and the emitter through a Josephson junction.
    In the original BOT the base has a normal tunnel junction (NIS), whereas in the superconducting version this is replaced by a Josephson junction.
    Capacitances of the junctions are shown explicitly. We assume current bias at the base and voltage bias at the collector.
(b): The central charge-tunneling processes in the original BOT. The current from the collector drives the quasicharge as
     $\dot q=I_{\rm C}$.
     In a Bragg reflection (blue arrows) a Cooper-pair gets resonantly transferred from the emitter to the island while
     the system passes a resonance point $q=e,3e,\ldots$,
     whereas in a Zener tunneling (black arrow) the charge is accumulated on the island (capacitors) and increases its voltage until it reaches $V_{\rm C}$.
     Ideally, this static situation can be escaped only by a quasiparticle tunneling to the base (red arrow), providing base-current amplification
     when followed by repeated Bragg reflections.
}
\label{fig:bot}
\end{figure}

 Superconducting microelectronics with nonlinear components, such as the Josephson junction (JJ),
 offer a vast arena for elementary quantum mechanics and its applications.
 Particularly, JJs exhibit energy-band phenomena similar to solid-state physics, such as the Bragg reflection
 and Zener tunneling~\cite{Averin,Likharev,SchoenZaikin,Kuzmin1991,Haviland1991}.
 The Bloch-oscillating transistor (BOT) is a creative application of these phenomena~\cite{BOT}.
 It is based on triggering long-lasting periodic motion to a (fictitious) particle traveling along an energy band, by a short perturbation to its momentum.
 In its microelectronic realization~\cite{BOT,BOTTheory,LindellAPL2005,LindellJLTP2009,Sarkar2013,SarkarDBOT},
 an incoherent quasiparticle tunneling between the normal-state base lead and superconducting island [see Fig.~\ref{fig:bot}(a)],
 triggers repeated resonant Cooper-pair tunnelings
 through the emitter JJ to the island (Bragg reflections) and charge flow to the collector.
 This can be used for low-noise current amplification, especially for intermediate source impedances~\cite{BOT,BOTTheory,LindellAPL2005,LindellJLTP2009}.
 The research of Bloch~\cite{BOT,BOTTheory, LindellAPL2005, LindellJLTP2009, Sarkar2013,SarkarDBOT, Watanabe2001,Watanabe2003, Corlevi2006,Zorin2006,Nguyen2007,Weissl2014} and generally single-charge oscillations~\cite{Pistolesi2012} in microelectronic circuits
 is also motivated by the possible applicability to metrology~\cite{Metrology}, quantum computing~\cite{Makhlin},
 and noise detection~\cite{LindellPRL2004}.

 In this article, we study experimentally and theoretically a  Bloch-oscillating transistor that is based only on Cooper-pair tunneling.
 A continuous current flow to the superconducting island is established by voltage biasing a high-resistance thin-film resistor
 in the immediate vicinity of the transistor island.
 This drives the quasimomentum of the system, corresponding to charge flow between the island and the collector.
 Connection to the emitter and the base is realized through superconducting tunnel junctions.
When biased below the superconducting gap, we find that  the periodic motion of the quasimomentum controls Cooper-pair transport across the emitter and the base junctions. Near two operation points the system can be made to switch between largely differing transport regimes by a small change in the base current. At high tunnel couplings the transport is found to bifurcate, so that the behavior becomes hysteretic at the two operation points.

 For a detailed understanding of the underlying physics, we establish a density-matrix description of
 energy-band dynamics induced by three factors: 1) the collector bias, 2) Cooper-pair tunneling,
 and 3) fluctuations and dissipation due to resistive parts of the bias circuitry.
 We find that at one side of an operation point,
 there exists a fragile balance between two base JJ Cooper-pair tunneling processes, occurring repeatedly to two opposite directions.
 A small change in the bias can lead to a situation where these counteracting processes do not anymore neutralize each other,
 and the system is forced to switch to a regime with lower transport and no counterflow. 
 A shifted balance can be used for strong amplification of the base current, which in this case is formed from Cooper pairs
 in contrast to single electrons in the traditional BOT structure.

 We present the results of this work in the following way. In Section~\ref{sec:BOT}, 
 we discuss in a simplified way how the traditional working principle of BOT changes when the BOT is made fully superconducting.
 In Section~\ref{sec:experiment} we go through the experimental details, and
 in Section~\ref{sec:theory} we introduce our theoretical model that accounts for the
 most important energy-band dynamics in the system.
 In Section~\ref{sec:results} we compare the experiments with numerical simulations and discuss in detail the physics behind
 the observed phenomena.
 Conclusions and outlook are given in Section~\ref{sec:Conclusions}.


\section{The Bloch-oscillating transistor}\label{sec:BOT}
\subsection{Original working principle}
The original working principle of the Bloch-oscillating transistor is depicted in Fig.~\ref{fig:bot}.
The voltage applied on the collector attracts charge carriers from the island to flow through the resistor $R$
to the collector. This in turn leads to a motion of the quasicharge $q$, that plays the role of the quasimomentum here.
The large collector resistance $R\gg R_Q\equiv h/4e^2$ guarantees small quantum fluctuations of the quasicharge~\cite{Averin,Likharev,IngoldNazarov},
and the system description in terms of energy-bands $E^n(q)$ with semiclassical $q$ is possible.
A charging energy that is comparable to the Josephson coupling, $E_C \sim E_{\rm J}$, where $E_C=e^2/2C$ and $C$ the capacitance of the island,
provides small band gaps between higher energy bands, also an essential property in the following.
Analogously to electronic states in solid, in the neighbourhood of avoided level crossings the system
can either perform a Bragg reflection, i.e.~stay on the same band, or Zener tunnel to the higher energy band.
Here, this corresponds to a tunneling of single Cooper-pair to the island, or absence of it, respectively~\cite{SchoenZaikin}.

Repeated Bragg reflections, called Bloch oscillations, allow charge to flow between the emitter and the collector, until
at some point a Zener tunneling occurs and the potential of the island starts to increase.
If further tunnelings are also inhibited,
due to small gaps in higher bands, charge flow through the
collector resistor stops only after the island voltage $\partial E^n(q)/ \partial q$ reaches the collector voltage.
If one neglects incoherent Cooper-pair tunneling through the emitter,
the system can be made to switch back to the lowest energy band only by letting the electrons to tunnel to the base lead through the NIS junction.
Therefore, if properly biased, single-electron tunneling events to the base can lead to multiple Cooper-pair tunnelings across the emitter,
and therefore to current amplification.
This is the original working principle of the Bloch-oscillating transistor, discussed more extensively in Refs.~\onlinecite{BOT,BOTTheory}.

\begin{figure}[bt]
\includegraphics[width=\linewidth]{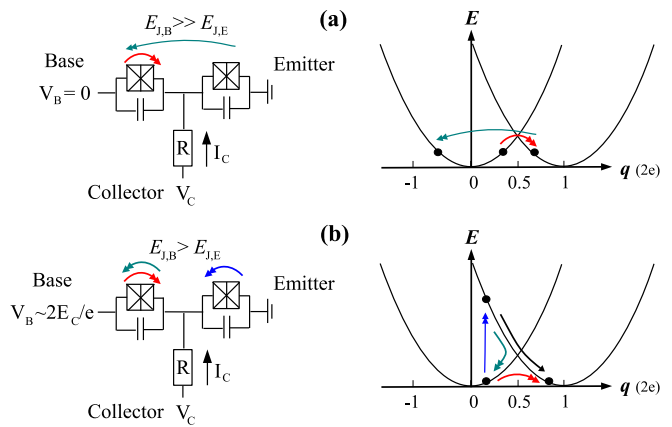}
\caption{Charge-transport processes in the fully superconducting BOT, that provide amplification of the base current.
(a): In the first bias point Cooper pairs can be made, with a small change in the base current,
to resonantly cotunnel from the emitter to the base, virtually over the island with no induced energy-band transition,
and after this to "counterflow" to the island, via a Bragg reflection.
Due to large $E_{\rm J,B}$, this provides higher collector current as for transport in the absence of the base.
(b): 
In the neighbourhood of the second operation point, two counterflowing components (Bloch oscillations
in the lowest band and "counter-dropping" of the quasicharge to the lowest band)
hold the voltages of the electrodes in favourable values for intensive resonant Cooper-pair across the emitter
(upward transition).
A small change in the base current can outbalance the situation and lead to only weak off-resonant Cooper pair transport.
}
\label{fig:processes}
\end{figure}

\subsection{Fully superconducting case}\label{sec:S-BOT_Short}

When the base lead is made superconducting, one could expect a similar transport scheme as in the original BOT, but with Cooper-pairs tunneling to the base.
We find this to be qualitatively true, the base current can be amplified through repeated Bloch-oscillations,
but through a more subtle tunneling procedure.
The difference stems from that, unlike in the case of quasiparticles,
the Cooper-pair tunneling occurs predominantly resonantly,
which means preference for a certain voltage-difference window.
Here, we observe two new amplifications schemes
based on resonant Cooper-pair tunneling across  both JJs.
Below we summarize the results in a simplified manner, and later in Section~\ref{sec:results} give a more quantitative analysis based
on our numerical simulations.

\subsubsection{Voltage trapping nearby $V_{\rm B}=0$}
In the amplification scheme based on a large Josephson energy at the base JJ, $E_{\rm J,B}\gg E_{\rm J,E}$,
the current brought to the base is used to trap the phase (time-integrated voltage) between the base and the emitter.
When trapped nearby $V_{\rm B}= 0$, the Cooper-pair coming from the emitter can resonantly cotunnel to the base,
and from there resonantly tunnel to the island. From the island it flows to the collector, as visualized in Fig.~~\ref{fig:processes}(a).
In the trapped state, the two tunneling processes across the base-junction occur in opposite directions.
At the limit where the two counterflowing current components can barely balance each other (tunable by $V_{\rm C}$),
a small change in the base current can overcome the maximal cotunneling transport between the emitter end the base,
and the base voltage $V_{\rm B}$ is forced to move closer to the collector voltage in order to balance the incoming and outgoing
current components. This leads to modified operation in which
Cooper pairs from the emitter get predominantly transported directly to the island,
but with a smaller frequency than in the state with $V_{\rm B}\sim 0$, due to Coulomb blockade of Cooper-pair tunneling.
 Such amplification is analogous to a classical current biased Josephson junction, where the phase difference (between the base and the emitter)
 can be switched to a running state with a small change in the feed current~\cite{Tinkham}.

\subsubsection{Voltage trapping nearby $V_{\rm B}=2E_C/e$}
In this article, the resonant Cooper-pair tunneling across the base JJ
is described as an avoided energy-level crossing, and the resonant tunneling across the emitter JJ
as a vertical energy-level transition.
In an emitter tunneling event, the voltage of the island decreases, and subsequently starts to be recharged through the collector current.
The momentarily lowered island voltage in turn induces resonant base tunneling.
In Fig.~\ref{fig:processes}(b), this means that after an upward transition (blue arrow), while sliding downwards in the excited energy-band
(recharging, black arrow),
a Cooper pair can jump resonantly across the base JJ in the neighbourhood of $q=e$ (turquoise arrow).
Since the net transport to the base is fixed, this "counterflow" decreases $V_{\rm B}$.
After the quasicharge has returned to the lowest energy-band, the change in $V_{\rm B}$ pushes the quasicharge rightwards
and induces an opposite-direction base tunneling, i.e.~a Bragg reflection (red arrow).
In a steady state, the base voltage settles into a position where these two processes compensate each other, usually slightly below $V_{\rm C}$.

Increasing the bias $V_{\rm C}$, and coming closer to $2eV_{\rm B}\approx 4E_C$,
the rate of these to counteracting processes increases,
as the resonance position of the upward transition starts to "hit" the propable position of the quasicharge
(that is in the neighbourhood of the bottom of the lowest energy band).
When further increasing $V_{\rm C}$, the base voltage is trapped to $2eV_{\rm B}\approx 4E_C$,
favourable for intensive upward transitions as well as increased Bragg reflections.
For certain $V_{\rm C}$, the probability for the (strongly driven) quasicharge
to be in a region of resonant upward transition, starts to decrease.
This in turn decreases Bloch oscillations, and the trapping of $V_{\rm B}$ starts to change to a rapid release.
The dynamics is nonlinear, and there is no guarantee that this process is continuous.
Indeed, a small change in  $V_{\rm C}$ (or base current) can lead to a rapid switch of $V_{\rm B}$
closer to $V_{\rm C}$.
This means off-resonance in Cooper-pair transport across the emitter and reduction of charge transport to the collector.

\begin{table}[b]
\begin{ruledtabular}
\begin{tabular}{ccccccc}
 \hline
 $R$ & $R_{\rm T,E}$ & $R_{\rm T,B}$ & $E_{\rm J,E}^{\rm max}$ & $E_{\rm J,E}^{\rm min}$ & $E_{\rm J,B}$ & $E_C$         \\
 \hline 41~k$\Omega$  &  4.5~k$\Omega$  &    10~k$\Omega$  &     140~$\mu$eV         &      $<5$~$\mu$eV          &   65~$\mu$eV  &   40~$\mu$eV     \\
 \hline
\end{tabular}
\end{ruledtabular}
\caption{Parameters of the studied sample. The values are based on the data shown in this article and additional independent two-terminal charge-transport measurements.
The emitter JJ is realized in the dc-SQUID geometry, having a tunable Josephson coupling energy $E_{\rm J,E}$.}\label{tab:parameters}
\end{table}

\begin{figure}[bt]
\includegraphics[width=0.9\linewidth]{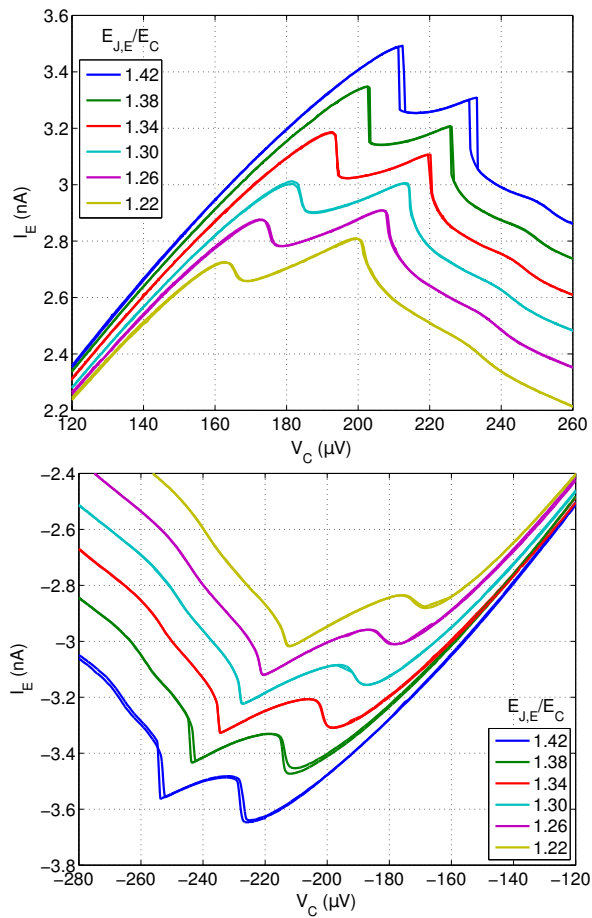}
\caption{
Measured emitter-collector $I_{\rm E}$-$V_{\rm C}$ curves for two polarities (signs) of the collector voltage (top and bottom)
for $I_{\rm B} = 200$~pA and for several values of $E_{\rm J,E}/E_C$.
The voltage is swept to two directions.
We observe a two-step structure, evolving to hysteresis with increasing $E_{\rm J,E}/E_C$.
The highest current gain value of 22 was observed just before the hysteresis.
The step-structure is surprisingly similar for the two polarities,
the major difference being a shift in the absolute values of the voltage and the current.
}
\label{fig:experiment1}
\end{figure}

\section{Experiment}\label{sec:experiment}
The studied sample consists of a single SIS-junction at the Aluminium base, a double SIS-junction in the dc-SQUID geometry at the emitter, and
of a thin-film Cromium resistor of 10~nm thickness at the collector, in a transistor-type structure connecting to a superconducting Aluminium island.
Sample parameters are given in Table~\ref{tab:parameters}.
The difference to earlier realizations of the Bloch-oscillating transistor~\cite{BOT,BOTTheory,LindellAPL2005,LindellJLTP2009,Sarkar2013,SarkarDBOT} is that there the system had an NIS-junction at the base,
while here we have an SIS junction, with the Josephson coupling energy, $E_{\rm J,B}$.

Measurements were done in a Nanoway PDR50 dilution refrigerator at a base temperature of 60~mK.
Additionally, a current bias to the base lead was established using a DC voltage source through a 1~G$\Omega$ resistor.
The base lead, in the vicinity of the JJ, provides a relatively stable voltage due to its large capacitance to ground.
Therefore, the base voltage can be assumed to be a slowly-moving variable in the time-scale of single-charging processes.
For a more detailed description of experiment see Ref.~\onlinecite{KorhonenThesis}.

The emitter-collector $I_{\rm E}$-$V_{\rm C}$ curves for two polarities of the collector current
(the normal and the inverted BOT-operating regions~\cite{BOT,BOTTheory,LindellAPL2005,LindellJLTP2009,Sarkar2013,SarkarDBOT}) are shown in Fig.~\ref{fig:experiment1}, for 200~pA base-current.
Compared to earlier NIS measurements, we observe the following unexpected features:
1) the appearance of a two-step structure, instead of a single-step,
2) robustness of the form of the step-structure under the change of the bias current or its polarity, and 3)
  bifurcation at voltages nearby $\Delta/e$ (see $E_{\rm J,E}/E_C=1.42$), well below the quasiparticle tunneling threshold $2\Delta/e$.
These features can be reproduced with our density-matrix simulations, essentially only based on
modeling resonant Cooper-pair-tunneling processes across the two junctions with simultaneous quasicharge motion
driven by the collector current.


\section{Theory}\label{sec:theory}

Our theoretical model is based on the description of the resistively-shunted current-biased JJ~\cite{Averin,Likharev}.
This system is equivalent to a voltage-biased JJ with a series resistor, that corresponds to the experimental realization.
In the limit $R\gg R_{\rm Q}=h/4e^2$ the current fluctuations due to the shunt resistor are small and a good starting point is the Hamiltonian
of the current-biased Josephon junction~\cite{Averin,Likharev},
\begin{equation}\label{eq:Hamiltonian}
H=\frac{Q^2}{2C}-E_{\rm J} \cos \varphi - \frac{\Phi_0}{2\pi}\varphi I.
\end{equation}
This is a sum of capacitive energy, Josephson coupling energy, and a linear tilt describing the current source, correspondingly.
The charge $Q$ at the junction capacitor $C$ and the phase difference across the Josephson junction $\varphi$
are conjugated variables, $[\varphi,Q]=2ei$.
The Hamiltonian is analogous to the one of an electron in a periodic lattice, with an externally applied electric field.
In this article, we use this Hamiltonian for the description of Cooper-pair tunneling across the base JJ,
whereas tunneling across the emitter JJ is treated perturbatively, see Section~\ref{sec:S-BOT}.

The charging energy of a single Cooper pair on the island is rather large, $E_{\rm J}/4E_C< 1$,
and therefore single charging effects are essential. Also $R\gg R_{\rm Q}$ supports squeezing of the charge fluctuations on the island.
Due to commutation relation between the charge and the phase, this means that quantum fluctuations of the phase are large.
Therefore, an alternative approach based on assuming a localized phase difference at a minimum of the tilted washboard potential~\cite{Tinkham}
[the last two terms in Hamiltonian~(\ref{eq:Hamiltonian})] cannot be used here.
However, at some points we use this concept for visualization of voltage-trapping between the base and emitter
due to resonant tunneling between the leads, characterized by a small charging energy and resistivity below the resistance quantum $R_{\rm Q}$.

\subsection{Density matrix}
Our basis for density-matrix simulations is the eigenstates of Hamiltonian~(\ref{eq:Hamiltonian}) when $I=0$.
They form $2e$-periodic energy bands $E^n(q)$ as a function of the quasicharge $q$\cite{Averin,Likharev,Prance,Flees,LindellPRB2003,Transmon}.
The quasicharge corresponds physically to the charge fed to the JJ by the external circuit.
We denote the energy bands by $n$ and introduce a dimensionless quasimomentum $k=q/2e$.
Following Ref.~\onlinecite{Likharev}, we assume that the density matrix
\begin{equation}
\rho=\sum_{nn'}\int dk\int dk'\rho^{nn'}(k,k')\vert n,k\rangle\langle n',k'\vert,
\end{equation}
has the diagonal form $\rho^{nn'}(k,k')=\rho^{nn'}(k)\delta(k-k')$. 
The static part of the current bias is treated exactly, as shown below, causing time dependent motion to the quasicharge~$q$
and coherent transitions between the energy bands.

\subsubsection{Density matrix equation of motion}
The coherent Hamiltonian, Eq.~(\ref{eq:Hamiltonian}), leads to the quantum Liouville equation
\begin{equation}\label{eq:Liouville}
\dot\rho=\frac{i}{\hbar}[H,\rho]=\frac{i}{\hbar}[H_0,\rho]+\frac{i}{\hbar}[H_I,\rho],
\end{equation}
where $H_I=-\Phi_0\varphi I/2\pi$. The commutator with $H_0$
produces phase factors in nondiagonal elements as we work in the eigenbasis of $H_0$. For evaluating
the commutator with $H_I$ one can use the expression
\begin{equation}
\varphi_{kk'}^{nn'}=i\delta_{nn'}\frac{\partial\delta(k-k')}{\partial k}+\varphi_{k}^{nn'}(1-\delta_{nn'})\delta(k-k'),
\end{equation}
where the matrix elements are given in the Appendix.
For an arbitrary density-matrix entry $\rho^{nn'}(k)$ this leads to the equation of motion
\begin{eqnarray}
\dot\rho^{nn'}(k)&=&-i\omega_k^{nn'}\rho^{nn'}(k)-\frac{I}{2e}\frac{\partial\rho^{nn'}(k)}{\partial k}\label{tiheys1}\\
&+&i\frac{I}{2e}\left[\sum_{n_i\neq n}\varphi^{nn_i}_k\rho^{n_in'}(k)-\sum_{n_i'\neq n'}\varphi^{n_i'n'}_k\rho^{nn_i'}(k)\right].\nonumber
\end{eqnarray}
Here $\omega_k^{nn'}=[E^n(k)-E^{n'}(k)]/\hbar$.
The first term in the right hand side of Eq.~(\ref{tiheys1}) corresponds
to evolution without the bias current, the second one describes the current-bias
and results in a time dependence $\dot q=I$, and the last
term describes interband (Zener) tunneling due to the bias current.

\subsection{Effect of the resistor}
The fluctuations in the feed current, due to a coupling to the shunt resistor, are taken into
account as a perturbation, producing both classical and quantum
fluctuations to the current.
In Hamiltonian of Eq.~(\ref{eq:Hamiltonian}), we add to the dc-bias $I$ a fluctuating component $\hat{I}$,
satisfying the quantum-mechanical equilibrium correlations
\begin{eqnarray}
D(\omega)&=&{\rm{Re}}\left\{\left(\frac{\Phi_0}{2\pi}\right)^2\lim_{s\rightarrow 0}\int_{0}^{\infty}dt\langle\hat I(t)\hat I(0)\rangle e^{i\omega t-st}\right\}\nonumber \\
&=&\frac{\hbar^2}{2\pi}\frac{R_Q}{R}\frac{\omega}{1-e^{-\hbar\omega/k_BT}}.
\end{eqnarray}
The induced noise is taken into account by additional terms in the density matrix equation of motion, using a Born-Markov approximation.
This, and the assumed diagonality of the density matrix in quasimomentum $k$, are good approximations for small fluctuations of the current,
$\alpha=R_Q/R\ll 1$, and for short environmental memory times $\hbar/k_{\rm B}T<RC$. 

\subsubsection{Resistor in the density matrix equation}

 The resulting Born-Markov equations~\cite{Leppakangas2009} give three kind of extra contributions
 on the right hand side of Eq.~(\ref{tiheys1}). The first one describes dissipative quasicharge dynamics
 with no interband tunneling,
\begin{eqnarray}
\dot\rho^{nn'}_{\delta I}(k)&=&\frac{\alpha k_BT}{2\pi\hbar}\frac{\partial^2\rho^{nn'}(k)}{\partial k^2}\label{tiheys3} \\
&+&\frac{\alpha}{4\pi\hbar}\frac{\partial}{\partial k}\left\{\rho^{nn'}(k)\left[\frac{\partial E^n(k)}{\partial k}+\frac{\partial E^{n'}(k)}{\partial k}\right]\right\}.\nonumber
\end{eqnarray}
The first term on the right hand side of Eq.~(\ref{tiheys3}) describes quasicharge diffusion due to thermal fluctuations, and the second term
classical charge flow across the parallel resistor. 
Therefore, these terms correspond to a semiclassical (overdamped)
motion of the quasicharge in an energy band, subject to classical fluctuations and dissipation,
but no interband transitions~\cite{Averin,Likharev}.

\begin{figure}[tb]
\includegraphics[width=0.94\linewidth]{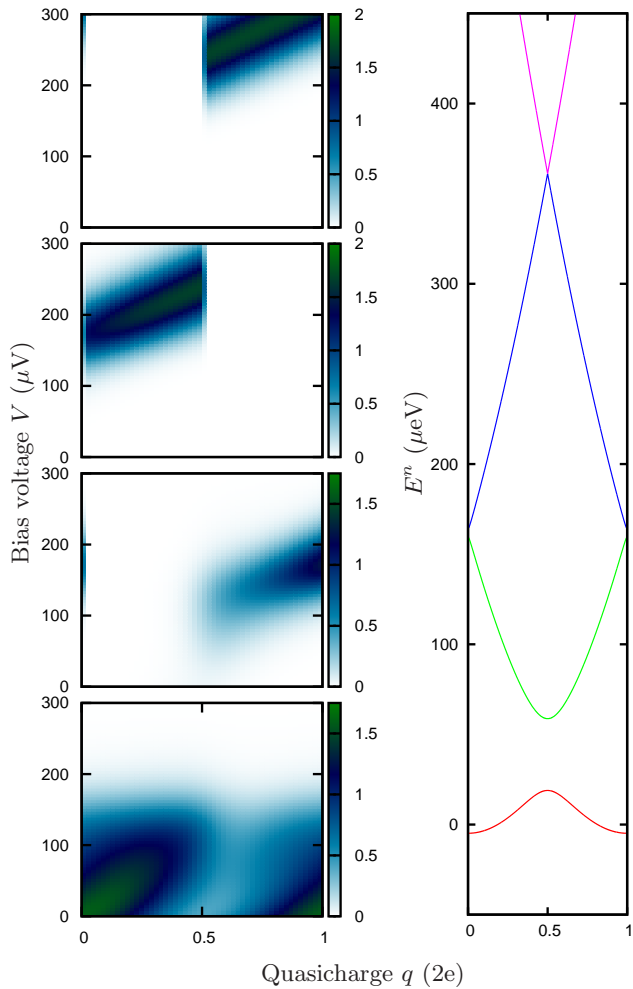}
\caption{ The steady-state probability density $\rho^{nn}(q/2e)$ (left) and the four lowest energy bands (right).
The density is always broadened due to presence of thermal fluctuations in the bias line,
typical width being $\delta V_{\rm J}=\sqrt{8E_C k_BT}/2e$~\cite{IngoldNazarov}. For small bias voltages $V$ the mean value of quasimomentum increases linearly, corresponding to linearly increasing junction voltage (see also Fig.~\ref{fig:currentbiasIV}).
In the region $V\approx 100$~$\mu$V the system performs fast Bloch oscillations in the lowest energy band,
that corresponds to low mean voltage across the JJ and maximum transport.
Above this the bias drives the system to higher energy bands, where Bragg reflections are rare due to negligible band gaps.
This means increased junction voltage and reduced transport.
In the parameter regime of the figure the first band-gap is close to $E_{\rm J}$, and the second band-gap to $E_{\rm J}^2/8E_C$.
We use here $E_{\rm J}=E_C=40~\mu$eV, $R=41$~k$\Omega$ and $T=200$~mK.
}
\label{fig:DM}
\end{figure}

\begin{figure}[tb]
\includegraphics[width=\linewidth]{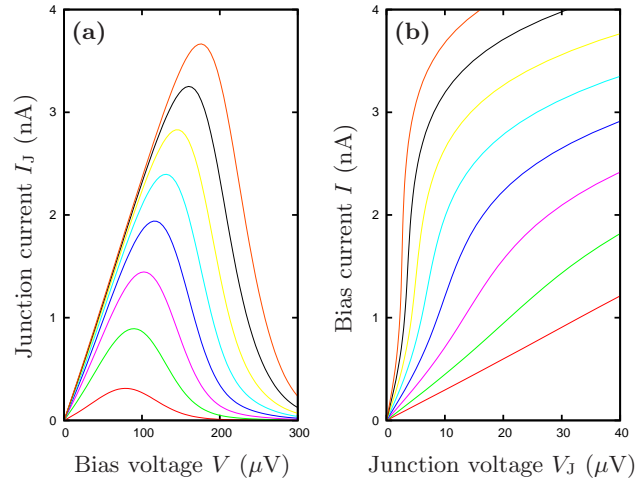}
\caption{
Current-voltage characteristics of a single current-biased JJ with $E_{\rm J}/E_C=1/4,1/2,\ldots ,7/4,2$, from the bottom to top. 
The other parameters are as in Fig.~\ref{fig:DM}.
In (a) we plot the junction current as a function of the bias voltage $V=IR$.
For small $E_{\rm J}$ we observe a thermally broadened peak around $V=e/C$,
reflecting the Coulomb blockade of Cooper-pair tunneling.
This result is familiar from the $P(E)$-theory~\cite{IngoldNazarov,DevoretPE}.
For higher values of $E_{\rm J}$ the curve is pushed against the resistor line $I=V/R$,
as a result of high conductance across the JJ due to intensive Bloch oscillations. At the onset of Zener tunneling the junction current starts to decrease.
In (b) we plot the junction voltage $V_{\rm J}$ as a function of the equivalent bias current $I$ [see Eq.~(\ref{eq:Equivalence})].
We see the emergence of the so-called "Bloch nose"\cite{SchoenZaikin,Corlevi2006}.
The first junction voltage bending point (top of the nose) corresponds to the onset of Bloch oscillations,
and the second bending to Zener tunneling.
}
\label{fig:currentbiasIV}
\end{figure}

The second contribution describes energy-band transitions with no quasicharge dynamics
\begin{eqnarray}
\dot\rho^{nn'}_{\rm T}(k)&=&\frac{1}{\hbar^2}\sum_{n_i\neq n,n_i'\neq n'}\rho^{n_in_i'}(k)\varphi^{nn_i}_k\varphi^{n_i'n'}_k(D^{n_in}_k+D^{n_i'n'}_k)\nonumber\\
&-&\frac{1}{\hbar^2}\sum_{n_i,n_v(n_i\neq n_v\neq n)}\rho^{n_i,n'}(k)\varphi^{n_vn_i}_k\varphi^{nn_v}_kD^{n_in_v}_k \\
&-&\frac{1}{\hbar^2}\sum_{n_i',n_v(n_i'\neq n_v\neq n')}\rho^{n,n_i'}(k)\varphi^{n_vn'}_k\varphi^{n_i'n_v}_kD^{n_i'n_v}_k. \nonumber
\label{tiheys4}
\end{eqnarray}
Here $D^{nn'}_k=D(\omega^{nn'}_k)$.
These terms account for, for example, inelastic Cooper-pair tunneling (spontaneous downwards energy-band transition)
with photon emission to the resistive environment~\cite{Golubev1992}.

Finally, the third contribution describes mixing of the two processes,
e.~g.~effect of dissipative shunt current to energy-band transitions,
\begin{eqnarray}\label{tiheys5}
&&\dot\rho^{nn'}_{\delta I+{\rm T}}(k)=-\frac{i}{\hbar^2}\frac{\partial}{\partial k}\sum_{n_i\neq n}\rho^{n_i,n'}(k)D^{n_in}_k\varphi^{nn_i}_k \\
&&+
\frac{i}{\hbar^2}\frac{\partial}{\partial k}\sum_{n_i'\neq n'}\rho^{n,n_i'}(k)D^{n_i'n'}_k\varphi^{n_i'n'}_k\nonumber\\
&+&i\sum_{n_i'\neq n'}\varphi^{n_i'n'}_k\left\{ \frac{\alpha k_BT}{2\pi\hbar}\frac{\partial\rho^{nn_i'}(k)}{\partial k}+\frac{\alpha}{4\pi\hbar}\rho^{nn_i'}(k) \tilde E^{n_i'n}(k) \right\}\nonumber\\
&-&i\sum_{n_i\neq n}\varphi^{nn_i}_k\left\{ \frac{\alpha k_BT}{2\pi\hbar}\frac{\partial\rho^{n_in'}(k)}{\partial k}+\frac{\alpha}{4\pi\hbar}\rho^{n_in'}(k)\tilde E^{n_in'}(k) \right\}. \nonumber
\end{eqnarray}
Here we have defined
\begin{equation}
\tilde E^{ij}=\left[\frac{\partial E^{i}(k)}{\partial k}+\frac{\partial E^{j}(k)}{\partial k}\right].
\end{equation}

Typical results for the steady-state density-matrix as a function of the bias is plotted in Fig.~\ref{fig:DM},
and for the corresponding junction $I-V$ characteristics in Fig.~\ref{fig:currentbiasIV}. When determining the junction current $I_{\rm J}$ from the
numerically calculated junction voltage $V_{\rm J}$, we have used the exact relations between the equivalent current biased
(with parallel resistor) and voltage biased (with series resistor) views of the same circuit,
\begin{equation}\label{eq:Equivalence}
I_{\rm J}=\frac{V-V_{\rm J}}{R} \,\,\,\,\,\,\, , \,\,\,\,\,\,\,\,\, CV_{\rm J}=\left\langle Q \right\rangle.
\end{equation}
Here $V=IR$. The expectation value is calculated using the steady-state solution for the density matrix~$\rho$.

\subsection{Quasiparticle tunneling}
Analogously to dissipative terms due to coupling to the resistor,
we add quasiparticle tunneling to the simulations in the Born-Markov approximation.
Quasiparticles turn out not to have any major role in the analyzed features observed in the experiment.
The standard perturbative treatment, applicable to low transparency tunnel junctions,
is shown in the Appendix.

\subsection{Superconducting BOT: Including the second Josephson junction perturbatively}\label{sec:S-BOT}

 For a description of a fully superconducting BOT, we attach a second JJ
 to the small island between the resistor and the original JJ. In this work we do this perturbatively.
 It turns out that it is the emitter junction, that is the best choice to be treated perturbatively,
 but similar physics emerges also from the opposite choice.
 This is consistent with the fact that the emitter has the smallest Josephson coupling energy in the analyzed situations.
 By treating one junction exactly, and the other perturbatively, we account for
 cotunneling events between the superconductors, which turns out to be the key for explaining experimental observations.

 The approach is well valid for $E_{\rm J,E}\ll E_{\rm J,B}$, but
 when the coupling energies are of the same magnitude, the treatment becomes more phenomenological.
 For a quantitative fit here, Cooper-pair tunneling across the two JJs should be treated on an equal footing,
 for example as in Refs.~\onlinecite{Brink1991Z,Brink1991,Leppakangas2008}, where also emitter tunneling
 contributes to the energy-band structure.
 However, our relatively simple theoretical model explains the main features seen in the experiments also in this
 parameter range.

  Treating the base junction exactly we use the formulas above, identifying $V=V_{\rm C}-V_{\rm B}$.
  We also have to renormalize the island charge $Q$ in the Hamiltonian of Eq.~(\ref{eq:Hamiltonian}), see the Appendix for details.
  To describe the emitter Cooper-pair tunneling as a perturbation, we include the following transition rates between the energy bands
  formed by the base JJ
 (the full form of the Born-Markov master equation used in the calculations is given in the Appendix)
 \begin{eqnarray}
 \dot\rho^{ff}_{\rm J,E}(k)&=&\sum_{\pm i}\frac{E_{\rm J,E}^2}{\hbar} \left\vert \langle f,k\vert e^{\mp i\varphi}\vert i,k\rangle \right\vert ^2  \label{rate}\\
 &\times & P\left[ E^{i}(k)-E^{f}(k)\pm 2eV_{\rm B}\right] \rho^{ii}(k)  \nonumber.
 \end{eqnarray}
 Here $+$ ($-$) corresponds to forward (backward) tunneling,  and the function  $P(E)$ accounts for energy conservation.
 Using Fermi's golden rule one gets $P(E)=\delta(E)$. This reproduces the observed double-peak structure,
 but with a very strong hysteresis that is not observed in the experiment. To account for this
 we have to add a rather large broadening to this function.
 Still, we want the $P(E)$-function to be peaked around $E=0$ and to satisfy detailed balance,
\begin{equation}
 P(-E)=e^{-\beta E}P(E).
\end{equation}
 The broadening that fits to experiments is of the order of voltage broadening on the island due to thermal fluctuations and Bloch oscillations.
 It is also similar as the broadening of $I-V$ characteristics measured between the emitter and the base with
the collector left floating.

The current across the emitter JJ can now be calculated as
\begin{gather}
I_{\rm E}=2e \int dk \sum_{n_i,n_i',n}{\rm{Re}}\left[\rho^{n_in_i'}(k)\Gamma^{n_i\rightarrow n}_{n_i'\rightarrow n}(k)\right],
\end{gather}
where $\rho$ is the steady-state density-matrix and
$\Gamma$ the Born-Markov transition-rate tensor due to emitter Cooper pair tunneling.
Positive direction tunneling is summed with positive signs and the negative direction terms with negative signs.
The quasiparticle current across the emitter is calculated similarly.
Finally, the current from the collector to the island can be deduced to be
\begin{gather}
I_{\rm C}=\frac{1}{R}\left( V-\frac{\langle Q\rangle }{C}\right).
\end{gather}
Here $C$ is the total capacitance of the island.
The base current can be deduced from the demand of current conservation
\begin{equation}
I_{\rm E}=I_{\rm C}+I_{\rm B}.
\end{equation}
Below we use this model for a detailed analysis of the experimental results.

\begin{figure}[tb]
\includegraphics[width=0.95\linewidth]{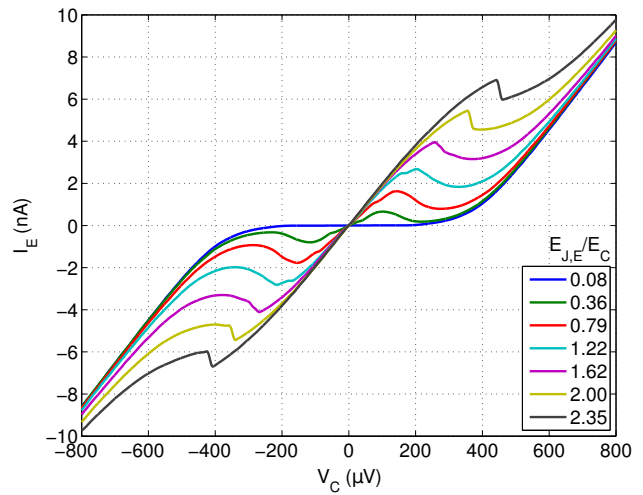}
\caption{
The overall form of experimental $I_{\rm E}-V_{\rm C}$ characteristics for $I_{\rm B}=0$ and several values of $E_{\rm J,E}/E_C$.
The collector voltage is swept from left to right.
For lowest values of $E_{\rm J,E}$ we observe a "$P(E)$-peak" at $V_{\rm C}\approx 80$~$\mu$V, and
for higher values  a behaviour close to $I=V_{\rm C}/R$.
Here, steps appear at voltages below the quasiparticle tunneling threshold $V_{\rm C}\sim 400$~$\mu$V,
evolving into hysteresis for highest values of $E_{\rm J,E}$.
 The sample parameters are given in Table~\ref{tab:parameters}.
}
\label{fig:OverallCurrent}
\end{figure}

\begin{figure*}[tb]
\includegraphics[width=0.85\linewidth]{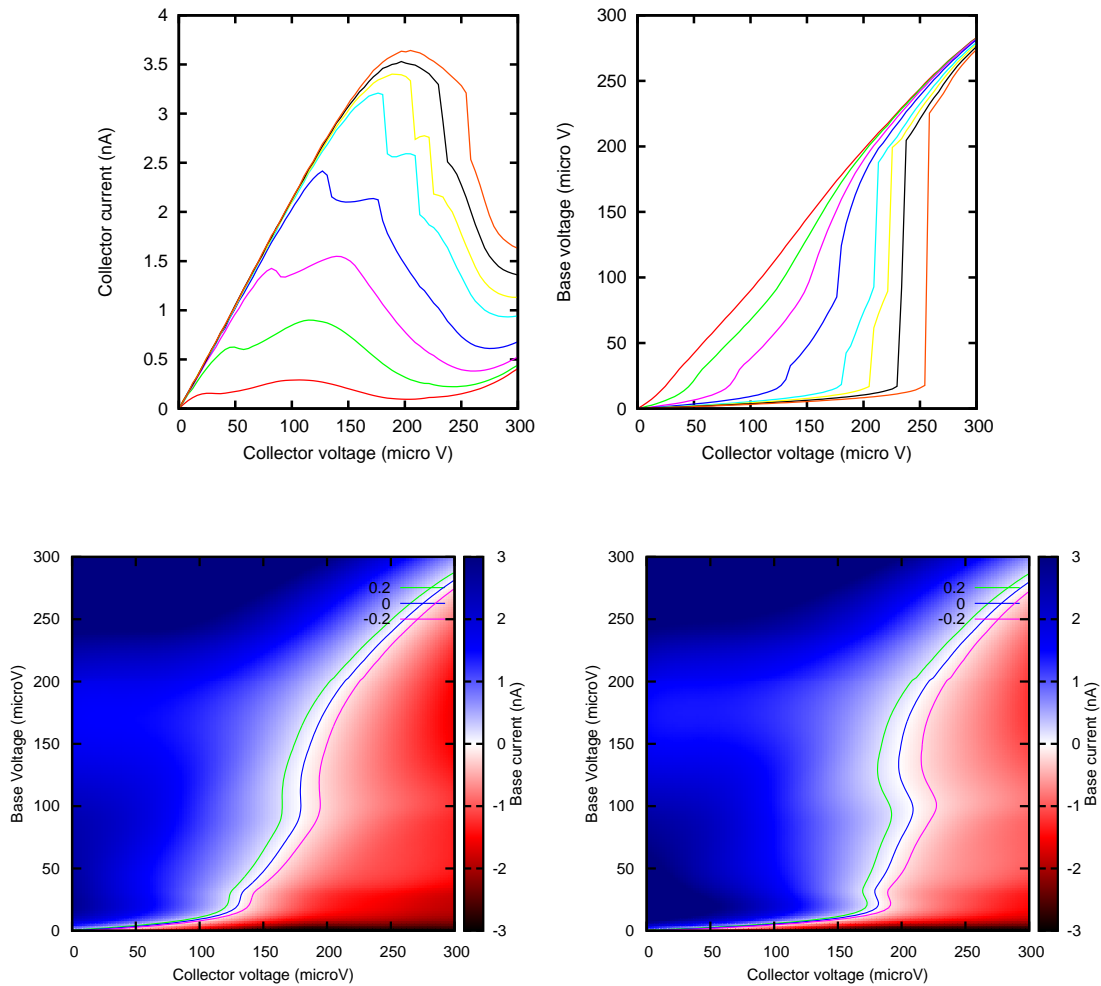}
\caption{
(Top left): Numerically calculated $I_{\rm E/C}-V_{\rm C}$ characteristics of a fully superconducting BOT, when $I_{\rm B}=0$
and voltage is swept from left to right.
    A double-step structure emerges on top of the normal $I-V$ peak of a current biased JJ
    when the Josephson coupling $E_{\rm J,E}$ of the emitter is increased.
(Top right): The base voltage in the same simulation, as a function of the collector voltage.
    For small $E_{\rm J,E}$, $V_{\rm B}$ follows $V_{\rm C}$, but by    
    increasing $E_{\rm J,E}$ the base voltage starts to trap nearby specific values, $V_{\rm B}=0$ or $V_{\rm B}=2E_C/e$,
    which support strong resonant charge transport between the emitter and the base.
(Bottom left): The base current as a function of base and collector voltages.
    The blue contour line ($I_{\rm B}=0$) corresponds  to the blue-line in (top). We also plot contour lines for $I_{\rm B}=0.2$~nA (green)
    and $I_{\rm B}=-0.2$~nA (pink).
    All the solutions are stable, since a small increment of the base voltage leads to a current away from the base.
    The general shape is robust against variations in the feed current.
(Bottom right): The same data for a slightly larger value of $E_{\rm J,E}$, the light-blue line in (top).
    Here, hysteresis appears in the neighbourhood of the studied bias points, and the picked stable solution depends to which direction
    the voltage is being swept.
    Parameters used in the simulation are the same as in Table~\ref{tab:parameters}, except $E_{\rm J,B}=70$~$\mu$eV and    
    $E_{\rm J,E}/E_{C}=0.25,0.5,0.75,1,1.25,1.375,1.5,1.625$,
    from the bottom to top, in the figure (top-left).
    We use here a Gaussian $P(E>0)$ with broadening 60~$\mu$eV and $P(E<0)=P(-E)e^{\beta E}$, see Section~\ref{sec:S-BOT}.
    We also use a (normalized) sub-gap density of states  $\rho(0<E<2\Delta)=0.3\times \left(E/2\Delta\right)^{10}$ with $\Delta=200$~$\mu$eV.
}
\label{fig:MainFigure}
\end{figure*}

\begin{figure}[tb]
\includegraphics[width=0.94\linewidth]{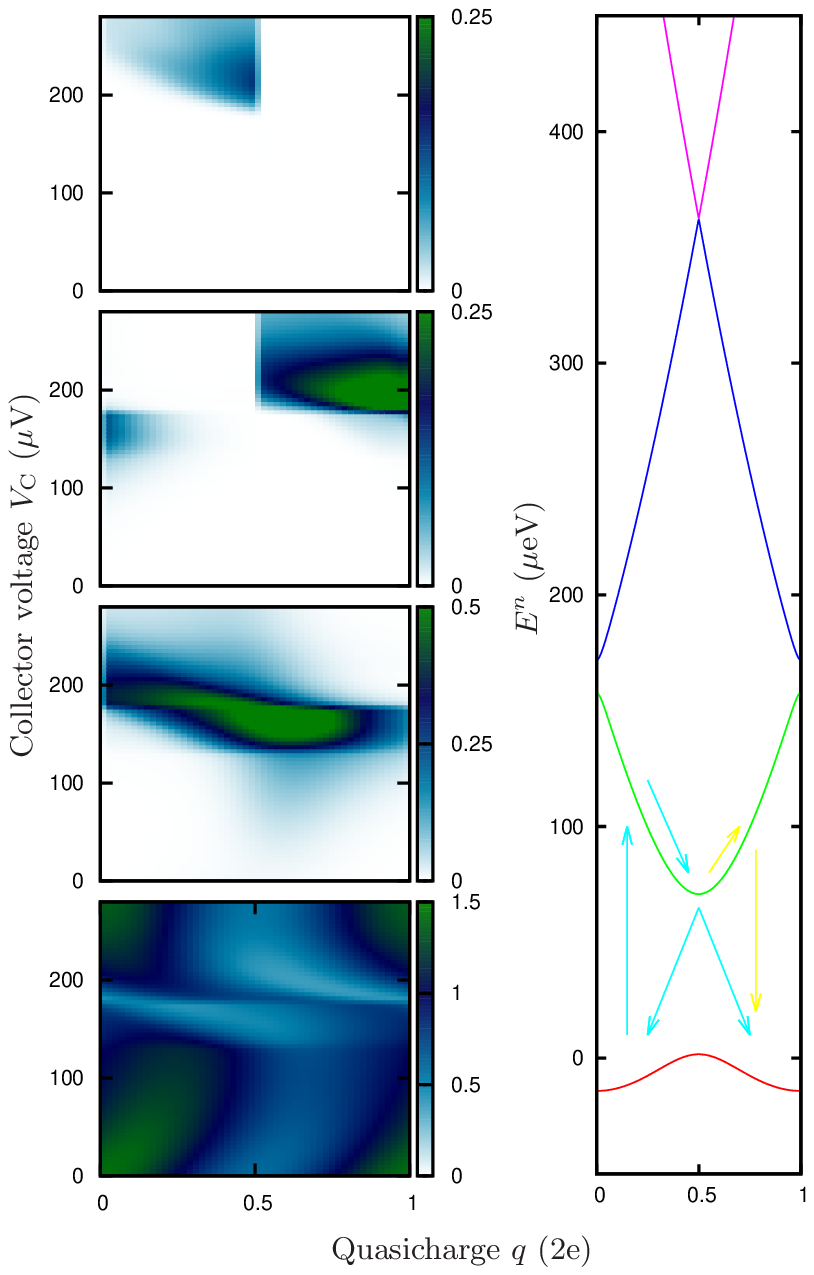}
\caption{
The steady-state probability density $\rho^{nn}(q/2e)$ (left) and the four lowest energy bands (right),
for a system as in Fig.~\ref{fig:MainFigure} with $E_{\rm J,E}/E_C=1$ and $I_{\rm B}=0$.
At low collector voltages ($<130$~$\mu$V), repeated Bragg reflections occur in the lowest energy band,
and can be identified from the tilt of the distribution to right.
For higher voltages, resonant Cooper-pair tunneling across the emitter increases the population of the first excited band.
The major part of the population is still in the lowest band.
Increasing $V_{\rm C}$ further moves the resonance position (the moving light-blue region in the lowest energy band)
of upward transitions leftwards.
For $V_{\rm C}> 180$~$\mu$V it passes the the bottom of the lowest band,
and population in higher bands reduces. After this the quasicharge populates mainly only the lowest energy band,
with weak temporal fluctuations triggered by off-resonant emitter tunneling.
}
\label{fig:DM2}
\end{figure}


\section{Results}\label{sec:results}
The general shape of experimental $I_{\rm E}-V_{\rm C}$ characteristics can be seen in Fig.~\ref{fig:OverallCurrent}.
If compared to previous BOT measurements~\cite{BOT,Sarkar2013}, several new features can be observed: the
disappearance of Coulomb blockade with increasing $E_{\rm J,E}$,
the appearance of two steps instead of single one for middle values of $E_{\rm J,E}$, and their bifurcation at voltages below
the quasiparticle tunneling threshold for higher values of $E_{\rm J,E}$.

\subsection{ Weak emitter tunneling limit }\label{sec:resultsA}
Typical numerical results for the current-voltage characteristics are shown in Fig.~\ref{fig:MainFigure}.
For small $E_{\rm J,E}$ we observe a Coulomb blockade at low voltages and
a "$P(E)$-peak" centered at $2eV_{\rm C}=4E_C$, as in experiments.
The transport is therefore similar as for a biased single JJ with $E_{\rm J}=E_{\rm J,E}\ll E_C$, Fig.~\ref{fig:currentbiasIV}(a).
This suggests that there is no significant interference coming from the base to the current between the emitter and the collector,
even though $E_{\rm J,B}\gg E_{\rm J,E}$.
The position of the peak gives the charging energy of the island, whereas the width is used to estimate the actual
temperature of the resistor, $T=200$~mK.

The similarity to single JJ transport turns out to be misleading, as Cooper-pair tunneling across the base JJ does occur, and will be important below.
Base tunneling occurs, as during recharging of the island, after an emitter Cooper-pair tunneling,
the system passes the point $q = e$.
In the energy-band picture, Fig.~\ref{fig:DM2}, this means that after an upward energy-band transition,
the quasicharge first slides downwards towards the energy-band minimum at $q=e$ (the island recharges). There, it
can drop either to the left or to the right-hand side of the maximum in the lowest energy-band (the light-blue arrows).
Dropping to the left-hand side means a backward base Cooper-pair tunneling if compared to Bragg reflections with increasing $q$,
see also Fig.~\ref{fig:processes}.
This  counterflow increases the voltage difference between the base and the collector, and becomes compensated
by forward-direction resonant base tunneling, which we call here as Bragg reflections.

\subsection{Disappearance of the Coulomb blockade}
For higher values of $E_{\rm J,E}$, the Coulomb blockade disappears at low voltages
and is replaced by a quasi-linear behaviour close to $I_{\rm C}=V_{\rm C}/R$.
Here, the base voltage  (the voltage between the base and the emitter)
is trapped close to a value $V_{\rm B}=0$, as opposed to the case of very small $E_{\rm J,E}$ where it stays close to $V_{\rm C}$,
see Fig.~\ref{fig:MainFigure} (top right).
This is a result of strong Cooper-pair cotunneling from the emitter to the base
and a balancing resonant Cooper-pair transport from the base to the island through Bragg reflections.

In the steady state density matrix, Fig.~\ref{fig:DM2}, the forced Bragg reflections can be identified from
the tilting of the thermally broadened probability density to the right. The cotunneling does not
cause energy-level transitions, and therefore there is only a weak population in the higher energy bands in this region.
Without the cotunneling between the emitter and the base [in Eq.~(\ref{rate}) neglecting transitions $i=f$],
the base voltage stays closer to the collector voltage, the remaining difference $V_C-V_B$ being related to the induced base Cooper-pair tunneling
during recharging of the island (or in general terms it is a result of a net current across the collector resistor).

\subsection{The first operation point ($I-V$ step)}
Before the first $I-V$ step the emitter-collector transport is enhanced due to
the possibility for a Cooper pair, coming from the emitter, to cotunnel to the base with a relatively high tunneling rate,
and subsequently to get transferred back to the island via a Bragg reflection (resonant tunneling) across the base junction.
Increasing the collector voltage increases both counterflowing components.
At a critical voltage the two components cannot anymore properly cancel each other, which
leads to charging of the base lead and increase of the base voltage,
as seen in Fig.~\ref{fig:MainFigure} (top right).
After the transition, Cooper pairs from the emitter get predominantly transported directly to the island, but with a smaller frequency.
This is because, on one hand, the weaker $E_{\rm J,E}$ has a trouble to overcome Coulomb blockade of Cooper-pair tunneling,
and on the other hand, we have lost another channel for Cooper-pair transport between the emitter and the collector.
In the numerically solved steady-state density-matrix, Fig.~\ref{fig:DM2}, we see the step as the emergence of population in
the second energy-band (starting at $V_{\rm C}\approx 130\ \mu$V),
due to enhanced upward energy-band transitions, describing direct tunneling from the emitter on the island.
This should be distinguised from Zener transitions
to higher energy-bands caused by  fast motion of the quasicharge,
as they   would demand considerable higher collector voltage and thus would be 
possible only for the highest plotted values of $E_{\rm J,E}$ in Fig.~\ref{fig:MainFigure}.

The numerically obtained critical value for the voltage between the emitter and the base, $V_{\rm B}$,
is independent of $E_{\rm J,E}$, as seen in Fig.~\ref{fig:MainFigure} (top right).
This is because the switching occurs nearby a fast decrease (or negativity) in the differential conductance of
charge transport between the base and the emitter at the ("supercurrent") region $V_{\rm B}\sim 0$ (not plotted).
In the used perturbative treatment, similar to the $P(E)$-theory~\cite{IngoldNazarov},
the form is independent of $E_{\rm J,E}$, and this is also physically the case as long as the cotunneling is limited by the emitter tunneling
coupling, i.e.~when $E_{\rm J,E}\ll E_{\rm J,B}$.

Noticeable is also that
the first step occurs well below the quasiparticle tunneling threshold, in accordance with experiments, supporting our interpretation that only Cooper-pair tunneling is involved,
as opposed to the case of normal BOT~\cite{BOT,BOTTheory,LindellAPL2005,LindellJLTP2009,Sarkar2013,SarkarDBOT}. Actually, the steps occur in the region of onset for
multiparticle tunneling~\cite{Wilkins,Shumeiko2007}, $eV=\Delta$.
However, according to our additional numerical simulations, this type of subgap tunneling has only a minor contribution
to the observed phenomena.

\subsection{The second operation point ($I-V$ step)}

The second step appears for strong emitter tunneling couplings, in Fig.~\ref{fig:MainFigure} for $E_{\rm J,E}>E_C$.
Resonant emitter Cooper-pair tunneling (upward transition) occurs only nearby a definite value of quasicharge. In Fig.~\ref{fig:DM2} it can be spotted as the leftwards moving light-blue region in the lowest energy band.
Simultaneously, the upward transitions lead also to a strong current component across the base JJ,
due to the backward-tunneling effect at $q=e$ (see the light-blue arrows in Fig.~\ref{fig:DM2} and the discussion above).
This increases the voltage difference $V_{\rm C}-V_{\rm B}$ until becomes compensated by
Bragg reflections in the lowest band.
The upward transitions become more frequent when there are better changes
for quasicharge to enter the resonance region (and stay there for a longer time).

In a stable solution, an increment of  $V_{\rm C}$ leads to an increment of $V_{\rm B}$, as visualized in Fig.~\ref{fig:MainFigure} (bottom).
Before the step, increasing $V_{\rm C}$ therefore moves the resonance position
of upward transitions to a region with more quasicharge population,
and increases the rate for the quasicharge probability to be pumped to the first excited band.
In the excited band, when sliding downwards, the quasicharge can avoid a transition to the lowest energy-band,
and instead get pushed to the right of $q=e$, due to high $V_{\rm C}-V_{\rm B}$.
This part of the distribution corresponds to island voltage equal to the collector voltage.
To get here, the quasicharge has performed a Bragg reflection in the first excited band, which corresponds to
backward base tunneling. From this region,
the quasicharge can drop to the lowest energy-band, for example, through inelastic (forward-directed) Cooper-pair tunneling across the emitter.
Such a route is visualized by the yellow lines in Fig.~\ref{fig:DM2}.

In the neighbourhood of resonant upward transition from the bottom of the lowest band, that is
for $2eV_{\rm B} \approx 4E_C$, we observe an opposite effect with increasing $V_{\rm C}$: the
previously occured latching of $V_{\rm B}$ changes to a rapid release, see Fig.~\ref{fig:MainFigure} (top right).
This reflects the fact that the time spend by quasicharge in the resonance region decreases.
Due to nonlinear nature of the phenomena involved,
there is no guarantee that this process is continuous, and for high enough $E_{\rm J,E}$
the reduction occurs as a discontinuous jump.
Quantitatively, an instability occurs if the lowest-band Bragg-reflection rate reduction via a small decrease $V_{\rm C}-V_{\rm B}\rightarrow V_{\rm C}-V_{\rm B}-\delta V_{\rm B}$ ($\delta V_{\rm B}>0$), is smaller than the counterflow reduction as a side product of upwards transitions with the base voltage $V_{\rm B}\rightarrow V_{\rm B} +\delta V_{\rm B}$.
Here an increment of the base voltage leads to a current towards the base, Fig.~\ref{fig:MainFigure} (bottom).

In the neighbourhood of the second step there appears also a new cotunneling process,
when the vertical transition hits the second (avoided) energy-level crossing.
This describes a higher-order tunneling process,
where not only a single Cooper-pair tunnels across the emitter JJ, but is accompanied by two Cooper-pairs across the base JJ~\cite{Brink1991Z,Brink1991,Leppakangas2008},
opening direct resonant transport between the emitter and the base.
However, according to our additional numerical simulations, where this effect is taken out
by removing all effects originating in the second energy-band splitting,
the new process does not explain the hysteresis, even though makes it stronger.

Finally, when the collector voltage is increased above the second step,
the base voltage switches close to the collector voltage, see Fig.~\ref{fig:MainFigure} (top right). Here
the Cooper-pair tunneling across the emitter (from the bottom of the lowest band) is inelastic.
The transport between the emitter and the collector, as well as across the base junction, reduce considerable.
Practically all of the probability density lies in the lowest band, symmetrically around $q=0$.
Weak temporal fluctuations are triggered by inelastic Cooper-pair tunneling across the emitter JJ.
This can then trigger further repeated resonant emitter tunneling events, now in higher energy bands, before quasicharge returns to the lowest band.

\subsection{Hysteresis and the relation to base-emitter fluctuations}

The bifurcation of the transport with increasing $E_{\rm J,E}$ is visualized in Fig.~\ref{fig:MainFigure} bottom.
Generally, the form of the base-voltage solution is very robust against variations in the base current,
basically just shifting the voltage-curve to the
corresponding direction, in agreement with the experiments, Fig.~\ref{fig:experiment1}.
In a stable solution, a small increase of the base voltage leads to positive base current (towards to the island). 
After a critical value of $E_{\rm J,E}$, this is not anymore true at the operation points, and the system becomes hysteretic,
having two different stable states for the same collector bias. The state is picked according to which direction the voltage $V_{\rm C}$ is swept.

In simulations the critical value of $E_{\rm J,E}$, at which the bifurcation occurs, is strongly related to fluctuations
in the voltage between the emitter and the base.
In theory, we would expect the base-emitter transport to be close to dissipationless supercurrent
[as for the Cooper-pair (or "Bloch") transistor~\cite{BlochTransistor1991,Tinkham}], due to low resistivity in the emitter and base leads.
This would lead to hysteresis that extends over the whole plotted voltage range,
resembling the case in a typical underdamped classical Josephson junction.
A possible high impedance environment would remove the hysteresis,
due to the absence of supercurrent that is replaced by incoherent cotunneling events across the two junctions~\cite{Zorin2003}.
However,  we observe no Coulomb blockade in transport between the emitter and the base,
supporting the presence of a noisy low-impedance environment, which also works against the hysteresis.
A relatively large broadening (see Section~\ref{sec:S-BOT}) have to be used to fit the experimental observations.
However, results from independent two-point measurements between the emitter and the base, where the collector is left floating, also imply
similar broadening. This data is characterized by the absence of supercurrent peak (negative differential conductance),
until $E_{\rm J,E}$ reaches a value that is close to the
beginning of hysteresis in the fully superconducting  BOT.
The source for the high noise is presumably the elevated temperature of the collector resistor caused by Joule heating.
Indeed, the used broadening is similar to thermal voltage fluctuations on the island induced by the collector.

\section{Conclusions and outlook}\label{sec:Conclusions}

 This article presents a detailed study
 of charge transport in a fully superconducting Bloch-oscillating transistor.
 We discovered new operation schemes, where Cooper-pair tunneling across the base Josephson junction
 is used to control quasicharge motion on the small superconducting island,
 and thereby  charge transport between the emitter and the collector.
 This occurred through trapping and detrapping of the base voltage,
 provided by the control of a counter-acting Cooper-pair transport across the base junction.   
 We found that the observed effects have a strong dependence on the Josephson coupling $E_{\rm J}$,
 which can be tuned (in real-time) by an externally applied magnetix field, allowing versatile control of the effect.

For applications, the important figures of merit are the current gain, the equivalent current noise at the input,
and the noise temperature. In our measurements, current gain up to 22 was observed, which is comparable to results of BOT devices with
NIS base junctions. The equivalent current noise above the 1/f corner frequency was found to be 3.5~fA/$\sqrt{\rm Hz}$ which is somewhat
higher than in regular BOTs having 1~fA/$\sqrt{\rm Hz}$ [\onlinecite{LindellAPL2005}]. The reason for this is most likely the relatively small collector
resistance $R$ [\onlinecite{BOTTheory}]. On the other hand, the small value of $R$ leads to reduced input impedance, which
results in a noise temperature of 0.2~K for the superconducting BOT. This noise temperature, in fact, wins over the results
on many regular BOTs (see, e.g.~Ref.~\onlinecite{LindellAPL2005}).
Indeed, the fully superconducting BOT has potential for a very low current noise, due to freeze-out of quasiparticles and
possibility for coherent charge transport. In principle, charge transport across the Josephson junction can be noiseless,
but noise will be inevitable due to the dissipation needed to avoid hysteresis.
Besides increasing the collector resistance and lowering its temperature for reduction of noise arising from the collector,
one can opt for an alternative device realization based on artificial low-noise high impedance environment, e.g. chain of Josephson junctions~\cite{Weissl2014}.
A significant noise reduction of the superconducting BOT device could be a key issue for breakthrough applications in metrology
or in quantum information processing.




\section*{Acknowledgments}
We thank M.~Marthaler, G.~Sch\"on, R.~Lindell, E.~Sonin,  M.~Hofheinz, and A.~Grimm for useful discussions.
Our work was supported by the Academy of Finland (contract 250280, LTQ CoE)
and by the Finnish Academy of Science and Letters (Vilho, Yrj\"o, and Kalle V\"ais\"al\"a Foundation).
The work benefitted from the use of the Aalto University Low Temperature Laboratory infrastructure.
AP is grateful to Finnish National Graduate School in Materials Physics for a scholarship.

\section*{Appendix: Theoretical methods}

\subsection*{Calculating matrix elements and energy bands}
To solve the density-matrix equation of motion, one needs to calculate matrix elements of $\varphi$. These can be represented as
\begin{gather}
\varphi_{kk'}^{nn'}=i\frac{\partial}{\partial k}\delta(k-k')\delta_{nn'}+\varphi^{nn'}_k\delta(k-k')(1-\delta_{nn'})\label{botkaava1}\\
\varphi^{nn'}_k=-i\left\langle n,k\left\vert\frac{\partial}{\partial k}\right\vert n',k\right\rangle.
\label{botkaava2}
\end{gather}
The first term on the right hand side of equation~(\ref{botkaava1}) describes intraband transitions whereas the second term interband transitions.
The element~(\ref{botkaava2}) is calculated between the $2\pi$-periodic parts of the wave functions and can be recast to the form
\begin{gather}
\varphi^{nn'}_k=-i\frac{\left\langle n,k\left\vert\frac{2eQ}{C_{\Sigma}}\right\vert n',k\right\rangle}{E_n(k)-E_{n'}(k)},\nonumber
\end{gather}
where $n'\neq n$. This is obtained by calculating the partial derivative of the Schr\"odinger equation
$H_{2\pi}(k)\vert n'\rangle=E_{n'}(k)\vert n'\rangle$ with respect to $k$
and taking matrix elements with respect to $\langle n\vert$.
The used Hamiltonian function is
\begin{gather}
H_{2\pi}(k)=\frac{(Q-2ek)^2}{2C}-E_J\cos\varphi,\nonumber
\end{gather}
and is solved with $2\pi$-periodic boundary conditions.

\subsection*{The second junction as perturbation}

A convenient Hamiltonian to start with for the case of fully superconducting BOT,
is the one of asymmetric Cooper-pair transistor~\cite{Leppakangas2008},
\begin{eqnarray}\label{botkaava0}
H&=&\frac{(Q+Q_0)^2}{2C_{\Sigma}}-E_{\rm J,B}\cos(\varphi) \nonumber\\
&& -E_{\rm J,E}\cos(\varphi-\varphi_{\rm E}) - Q_{\rm E}V_{\rm B}
\end{eqnarray}
Here $Q$ is the charge on the island and $\varphi$ a conjugated phase.
The capacitance of the island $C_{\Sigma}$ is the sum of the base and emitter JJ
capacitances, $C_{\Sigma}=C_{\rm B}+C_{\rm E}$, $E_{\rm J,B}$ and $E_{\rm J,E}$ are
the corresponding Josephson coupling energies. Charge operator $Q_{\rm E}$ and phase $\varphi_{\rm E}$ are conjugated variables.

The last two terms of Hamiltonian~(\ref{botkaava0}) describe tunneling across the emitter JJ and the corresponding
automatic work terms from the voltage sources.
A straightforward analysis shows that $(Q-Q_0)/C_{\Sigma}$ is the potential of the island. This allows us to
redefine the island charge as $Q\equiv Q-Q_0$. For description of the collector bias add the term $-\frac{\Phi_0}{2\pi}\varphi I $,
similarly as in Hamiltonian (\ref{eq:Hamiltonian}).

 We treat the base Cooper pair tunneling exactly, and
 the transition rates between the resulting energy bands due to emitter Cooper pair tunneling become,
 \begin{eqnarray}
 \Gamma_{i\rightarrow f}^{\pm}(k)&=&\frac{E_{\rm J,E}^2}{\hbar}\vert\langle n_f,k\vert e^{\mp i\varphi}\vert n_i,k\rangle\vert^2\nonumber\\
 &\times &P\left[E^{n_f}(k)-E^{n_i}(k)\mp 2eV_{\rm B}\right].\nonumber
 \end{eqnarray}
 The $P(E)$ function accounts for fluctuations in the voltage $V_{\rm B}$, and is basically treated as a source for broadening, see Sec.~\ref{sec:S-BOT}.
 The standard Born-Markov approximation~\cite{Leppakangas2008,Weiss} gives generalized transition rates,
 \begin{eqnarray}
 &&\Gamma_{b\rightarrow n}^{a\rightarrow m}(k)=\sum_{\pm }\frac{E_{\rm J,E}^2}{\hbar} \langle m,k\vert e^{\mp i\varphi}\vert a,k\rangle \left(\langle n,k\vert e^{\mp i\varphi}\vert b,k\rangle \right)^* \times \nonumber\\
 && \frac{1}{2} \{ P\left[E^{m}(k)-E^{a}(k)\pm 2eV_{\rm B}\right] + \nonumber \\
 && P\left[E^{n}(k)-E^{b}(k)\pm 2eV_{\rm B}\right] \}  \nonumber \\
 &&-\sum_{v\pm}\frac{E_{\rm J,E}^2}{2\hbar} \langle v,k\vert e^{\mp i\varphi}\vert a,k\rangle \left(\langle m,k\vert e^{\mp i\varphi}\vert v,k\rangle \right)^*\times\nonumber \\
 && P\left[E^{v}(k)-E^{a}(k)\pm 2eV_{\rm B}\right]\delta_{bn}\nonumber \\
  &&-\sum_{a,b,v\pm}\frac{E_{\rm J,E}^2}{2\hbar} \langle v,k\vert e^{\mp i\varphi}\vert b,k\rangle \left(\langle n,k\vert e^{\mp i\varphi}\vert v,k\rangle \right)^*\times\nonumber \\
 && P\left[E^{v}(k)-E^{b}(k)\pm 2eV_{\rm B}\right]\delta_{am}   .\nonumber
 \end{eqnarray}

\subsection*{Quasiparticle tunneling}
The operator describing the transitions on the island due to quasiparticle tunneling is,
\begin{gather}
\hat T=\sum_{Q}\vert Q-e\rangle\langle Q\vert. \nonumber
\end{gather}
In the case of the emitter JJ the quasiparticle tunneling and simultaneous transition $\vert n_i,k_i\rangle\rightarrow\vert n,k_f\rangle$ releases/absorbs the energy $\hbar\omega^{n_i}(k_i)-\hbar\omega^{n_f}(k_f)-eV_B$. Using the notation
\begin{gather}
\langle n',k'\vert T\vert n,k\rangle=T_{n'n}(k')\delta(k'-k+\frac{1}{2}) \nonumber \\
\langle n',k'\vert T^{\dagger}\vert n,k\rangle=T_{nn'}^{*}(k)\delta(k'-k-\frac{1}{2}), \nonumber
\end{gather}
one can calculate the terms originating in the positive-direction tunneling ($\hat T$-operator)
\begin{gather}
\dot\rho^{nn'}_{\rm{qp}}(k)=\sum_{n_i,n_i'}\rho^{n_in_i'}\left(k+\frac{1}{2},k+\frac{1}{2}\right)T_{nn_i}(k)T_{n'n_i'}^*(k)\nonumber\\
\times \{ D_q\left[\omega^{n_i}\left(k+\frac{1}{2}\right)-\omega^{n_f}(k)-\frac{eV_{\rm B}}{\hbar}\right]\nonumber\\
+D_q\left[\omega^{n_i'}\left(k+\frac{1}{2}\right)-\omega^{n_f'}(k)-\frac{eV_{\rm B}}{\hbar}\right] \} . \nonumber
\end{gather}
Here $D_q(eV/\hbar)$ is the quasiparticle tunneling rate across the corresponding voltage-biased single Josephson junction, and includes possible sub-gap density of states~\cite{Leppakangas2008,Tinkham}.
We have also contributions
\begin{gather}
\dot\rho^{nn'}_{\rm{qp}}(k)=-\sum_{n_i,n_v}\rho^{n_in'}\left(k\right)T_{n_vn_i}\left(k+\frac{1}{2}\right)\times \nonumber\\
 T_{n_vn}^*\left(k+\frac{1}{2}\right) D_q\left[\omega^{n_i}(k)-\omega^{n_v}\left(k+\frac{1}{2}\right)-\frac{eV_{\rm B}}{\hbar}\right] \nonumber ,
\end{gather}
and the mirror processes
\begin{gather}
\dot\rho^{nn'}_{\rm{qp}}(k)=-\sum_{n_i',n_v}\rho^{n,n_i'}\left(k\right)T_{n_vn_i'}^*\left(k+\frac{1}{2}\right) \times\nonumber\\
T_{n_vn'}\left(k+\frac{1}{2}\right) D_q\left[\omega^{n_i'}(k)-\omega^{n_v}\left(k+\frac{1}{2}\right)-\frac{eV_{\rm B}}{\hbar}\right]. \nonumber
\end{gather}
The opposite direction tunneling is obtained by the change $T\rightarrow T^{\dagger}$ and $eV_{\rm B}\rightarrow -eV_{\rm B}$. For the tunneling
across the base JJ the automatic work term $\pm eV_{\rm B}$ is set to zero.
We have taken the usual assumption that the imaginary parts of the transition rates are small compared to the energy-level differences,
which can be made for low-transparency tunnel junctions.

\subsection*{Discretizing the quasicharge space}

For numerical simulations the quasicharge, and derivatives with respect to it, have to be discretized. We do this
in a symmetric manner,
\begin{eqnarray}
\frac{\partial\rho}{\partial k} & \rightarrow & \frac{n}{2}\left (\rho_{k+1/n}-\rho_{k-1/n}\right) \nonumber \\
\frac{\partial^2\rho}{\partial k^2} & \rightarrow & n\left(\rho_{k+1/n}-2\rho_{k}+\rho_{k-1/n}\right). \nonumber
\end{eqnarray}
Here $n$ is the number of points in the discretized quasimomentum space, $k\in \{ 0,1/n,2/n\ldots 1-1/n \}$,
and $\rho_k$ the corresponding value $\rho(k)$. The derivatives of the energies have to
discretized carefully, in order to balance transition and decay rates of the quasicharge states.
For example, the definitions
\begin{eqnarray}
\frac{\partial E}{\partial k} & \rightarrow & n\left(E_{k+1/n}-E_{k} \right) \nonumber \\
\frac{\partial^2E}{\partial k^2} & \rightarrow & n\left( E_{k+2/n}-E_{k+1/n}-E_{k}+E_{k-1/n}\right),  \nonumber
\end{eqnarray}
lead to this property and preserve the trace of the density matrix. As our equation is not of Lindblad type,
the positiveness of the density matrix is not guaranteed either, but according to our
numerical simulations it holds for temperatures observed in the experiment.
For parameters as in the experiment, one runs into problems with positiveness for temperatures below $T\sim 50$~mK.

\subsection*{Numerical modeling}

The steady-state distribution of all the quasicharge dynamics is a solution of the discretized matrix-equation.
The equation has been greatly simplified by considering only diagonal states in $q$.
Still, it is important to discretize the quasicharge space much denser than typical changes occur in the corresponding energy-eigenstates. For small band-splittings this becomes a problem as changes occur rapidly nearby $k=0$ or $k=1/2$.
Therefore, in many cases the calculation is still quite challenging and needs to be solved with an appropriate method.

A numerically effective algorithm takes into account that only transitions between the nearby states of the quasicharge exist. This results in a block-tridiagonal matrix-equation for the steady state of the system, which has then its own effective inversion algorithm (Thomas algorithm). The upper-right and the lower-left corners of the matrix equation have still nonvanishing elements, which are not present in the block-tridiagonal form, but their effect can be included by using the Woodbury matrix identity.



\begin{thebibliography}{48}


\bibitem{Averin}
D. Averin, A. B. Zorin, and K. K. Likharev,
Sov. Phys. JETP {\bf 61}, 407 (1985).

\bibitem{Likharev}
K. K. Likharev and A. B. Zorin,
J. Low Temp. Phys. {\bf 59}, 347 (1985).

\bibitem{SchoenZaikin}
G. Sch\"on and A. D. Zaikin,
Phys. Rep. {\bf 198}, 237 (1990).


\bibitem{Kuzmin1991}
L. S. Kuzmin and D. B. Haviland,
Phys. Rev. Lett. {\bf 67}, 2890 (1991).

\bibitem{Haviland1991}
D. B. Haviland, L. S. Kuzmin, P. Delsing, K. K. Likharev, and T. Claeson, Z. Phys. B 85, 339 (1991)

\bibitem{BOT}
J. Delahaye, J. Hassel, R. Lindell, M. Sillanp\"a\"a, M. Paalanen, H. Sepp\"a, and P. Hakonen
Science {\bf 299}, 1045 (2003).

\bibitem{BOTTheory}
J. Hassel and H. Sepp\"a, J. Appl. Phys. {\bf 97}, 023904 (2005)


\bibitem{LindellAPL2005}
R. Lindell and P. Hakonen, Appl. Phys. Lett. {\bf 86}, 173507 (2005).


\bibitem{LindellJLTP2009}
R. Lindell, L. Korhonen, A. Puska, and P. Hakonen, J. Low. Temp. Phys {\bf 157}, 6 (2009)








\bibitem{Sarkar2013}
J. Sarkar, A. Puska, J. Hassel, and P. J. Hakonen, Phys. Rev. B, {\bf 87}, 224514 (2013).

\bibitem{SarkarDBOT}
J. Sarkar, A. Puska, J. Hassel, and P. J. Hakonen,  Supercond. Sci. Technol. {\bf 26}, 065009 (2013).



\bibitem{Watanabe2001}
M. Watanabe and D. B. Haviland, Phys. Rev. Lett. {\bf 86}, 5120 (2001).

\bibitem{Watanabe2003}
M. Watanabe and D. B. Haviland, Phys. Rev. B {\bf 67} 094505 (2003).

\bibitem{Corlevi2006}
S. Corlevi, W. Guichard, F. W. J. Hekking, and D. B. Haviland, Phys. Rev. Lett. {\bf 97}, 096802 (2006).

\bibitem{Zorin2006}
A. B. Zorin, Phys. Rev. Lett. {\bf 96}, 167001 (2006).

\bibitem{Nguyen2007}
F. Nguyen, N. Boulant, G. Ithier, P. Bertet,
H. Pothier, D. Vion, and D. Esteve, Phys. Rev. Lett. {\bf 99}, 187005 (2007).







\bibitem{Weissl2014}
T. Weissl, G. Rastelli, I. Matei, I. M. Pop,  O. Buisson, F. W. J. Hekking, and W. Guichard. Arxiv:1407.7728

\bibitem{Pistolesi2012}
C.~Negri and F.~Pistolesi, Phys.~Rev.~B {\bf 85}, 115416 (2012).

\bibitem{Metrology}
P. Francois, A. Bounouh, L. Devoille, N. Feltin, O. Thevenot, and G. Trapon,
Comptes Rendus Physique {\bf 5}, 857 (2004).

\bibitem{Makhlin}
Yu. Makhlin, G. Sch\"on, and A. Shnirman, Rev. Mod. Phys. {\bf 73}, 357 (2001).

\bibitem{LindellPRL2004}
R. K. Lindell, J. Delahaye, M. A. Sillanp\"a\"a, T. T. Heikkil\"a, E. B. Sonin, and P. Hakonen, Phys. Rev. Lett. {\bf 93} 197001 (2004).


\bibitem{Tinkham}
M. Tinkham, {\it Introduction to superconductivity}, 2nd ed. (Dover publications, 2004).


\bibitem{IngoldNazarov}
G.-L. Ingold and Yu.V. Nazarov, in {\it Single Charge Tunneling}, edited
by H. Grabert and M. H. Devoret (Plenum, New York, 1992), p. 21.


\bibitem{Brink1991Z}
A. Maassen van den Brink, A. A. Odintsov, P. A. Bobbert, and G. Sch\"on, Z. Phys. B: Condens. Matter {\bf 85}, 459 (1991).

\bibitem{Brink1991}
A. Maassen van den Brink, G. Sch\"on, and L. J. Geerligs
Phys. Rev. Lett. {\bf 67}, 3030 (1991).


\bibitem{Leppakangas2008}
J. Lepp\"akangas and E. Thuneberg, Phys. Rev. B {\bf 78}, 144518 (2008).

\bibitem{KorhonenThesis}
L. Korhonen, Master's thesis, Helsinki University of Technology, 2009.


\bibitem{Prance}
R. J. Prance, H. Prance, T. P. Spiller, T. D. Clark, Phys. Lett. A {\bf 166}, 419 (1992).

\bibitem{Flees}
D. J. Flees, S. Han, J. E. Lukens, Phys. Rev. Lett. {\bf 78}, 4817 (1997).

\bibitem{LindellPRB2003}
R. Lindell, J. Penttil\"a, M. Sillanp\"a\"a, and P. Hakonen, Phys. Rev. B {\bf 68}, 052506 (2003).


\bibitem{Transmon}
J. A. Schreier, A. A. Houck, Jens Koch, D. I. Schuster, B. R. Johnson, J. M. Chow, J. M. Gambetta, J. Majer, L. Frunzio, M. H. Devoret, S. M. Girvin, and R. J. Schoelkopf, Phys. Rev. B {\bf 77}, 180502(R) (2008).

\bibitem{Leppakangas2009}
J. Lepp\"akangas and E. Thuneberg, J. Phys.: Conf. Ser. {\bf 150}, 022050 (2009).


\bibitem{Golubev1992}
D. Golubev and A. D. Zaikin, Phys Lett. A {\bf 164}, 337 (1992).


\bibitem{DevoretPE}
M. H. Devoret, D. Esteve, H. Grabert, G-L. Ingold, H. Pothier, and C. Urbina, Phys. Rev. Lett. {\bf 64}, 1824 (1990).

\bibitem{Wilkins}
J. W. Wilkins, in {\it Tunneling Phenomena in Solids}, edited by E.
Burnstein and S. Lundqvist (Plenum, New York, 1969).

\bibitem{Shumeiko2007}
E. V. Bezuglyi, A. S. Vasenko, E. N. Bratus, V. S. Shumeiko, and G. Wendin,
Supercond. Sci. Technol. {\bf 20}, 529 (2007)

\bibitem{BlochTransistor1991}
D. V. Averin and K. K. Likharev, in {\it Mesoscopic Phenomena in Solids}, edited by B. L. Altshuler, P. A. Lee, and R. A. Webb
(Elsevier, Amsterdam, 1991).


\bibitem{Zorin2003}
S. V. Lotkhov, S. A. Bogoslovsky, A. B. Zorin, and J. Niemeyer, Phys. Rev. Lett. {\bf 91}, 197002  (2003)

\bibitem{Weiss}
U. Weiss, {\it Quantum Dissipative Systems}, 2nd ed. (World Scientific, Singapore, 1999).

































\end{thebibliography}
\end{document}